\newcommand{\x}{\hat{x}}
\newcommand{\xih}{\hat{\xi}}
\newcommand{\Q}{\hat{Q}}
\newcommand{\Qb}{\hat{\bar{Q}}}

\documentclass{article}
\usepackage{amsmath}
\usepackage{amsfonts}
\usepackage{amssymb}
\usepackage{hyperref}
\usepackage{amsmath,epsfig}
\begin{document}

\title{\textbf{Spinorial Snyder and Yang Models From Superalgebras And Noncommutative Quantum Superspaces}}
\author{\textbf{Jerzy Lukierski}, \textbf{Mariusz Woronowicz} \vspace{12pt} 
\\
Institute for Theoretical Physics, University of Wroclaw,\\
pl. Maxa Borna 9, 50-205 Wroclaw, Poland }

\maketitle
\date{}

\begin{abstract}
The relativistic Lorentz-covariant quantum space-times obtained by Snyder can be described by the coset generators of (anti) de-Sitter algebras. Similarly, the Lorentz-covariant quantum phase spaces introduced by Yang, which contain additionally quantum curved fourmomenta and quantum-deformed relativistic Heisenberg algebra,
 can be defined by suitably chosen coset generators of conformal algebras. We extend such algebraic construction to the respective superalgebras, which provide quantum Lorentz-covariant superspaces (SUSY Snyder model) and 
 indicate also how to obtain the quantum relativistic phase superspaces (SUSY Yang model). In last Section we recall briefly other ways of deriving quantum phase (super)spaces and we compare the spinorial Snyder type models defining bosonic or fermionic quantum-deformed spinors.

\end{abstract}

\numberwithin{equation}{section}

\section{Introduction}

In order to obtain quantum gravity models which reconcile two basic theories in physics, namely general relativity (GR) and quantum mechanics (QM), it appears desirable to introduce noncommutative (NC) quantum space-times (see e.g. \cite{1}-\cite{5}) and look for also the quantum deformations of quantum-mechanical relativistic phase space algebra with quantum NC fourmomenta (see e.g. \cite{6}-\cite{9}). Further, because the spinorial variables are even more fundamental that the vectorial ones (see e.g. quark model of hadrons or Penrose twistor theory) it is interesting to look for the algebraic systems providing quantum-deformed spinorial variables.

In this short paper we consider firstly various NC Lorentz-covariant  models
  of quantum space-times and quantum-deformed relativistic phase spaces,
 which originate from the ones proposed firstly  in 1947 by Snyder \cite{1} and Yang \cite{6}. 
 These historically first NC structures 
   of quantum space-times and quantum relativistic phase spaces are directly associated with the Lie 
     algebras describing D=4
     space-time symmetries which contain $D=4$ Lorentz algebra as its subalgebra. Further, these models will be naturally extended to supersymmetric ones, with supercharges promoted to fundamental quantum-deformed fermionic spinorial variables. 
     
      In standard $D=4$ Snyder  models
       one introduces NC space-time as described by the coset generators $O(4,1)/O(3,1)$ (dS Snyder model) or $O(3,2)/O(3,1)$ (AdS Snyder model)\footnote{We assume that the NC space-time coordinates
       and Lorentz $\hat{O}(3,1)$ generators are algebraically independent.
        Recently such models were studied under the name of extended Snyder models
        \cite{10},\cite{11}. We add that dS Snyder model is also called "Snyder model" and AdS Snyder model as "anti-Snyder model (see e.g.\cite{11a}).}. Each of these two models can be extended in two-fold way
        into Yang type models describing  quantum-deformed relativistic phase spaces
\begin{enumerate}
\item[i)]  dS Snyder model can be embedded into $O(5,1)$ or $O(4,2)$ Yang models, with quantum relativistic phase space variables $(\hat{x}_\mu, \hat{p}_\mu)$ as described by
  the generators of cosets  $O(5,1)/O(3,1)\otimes O(2)$ or $O(4,2)/O(3,1)\otimes O(1,1)$; 
\item[ii)] AdS Snyder model can be embedded into $O(4,2)$ or $O(3,3)$ Yang models with quantum relativistic phase space sector described by the coset generators of $O(4,2)/O(3,1)\otimes O(2)$ or $O(3,3)/O(3,1)\otimes O(1,1)$\footnote{$\hat{o}(4,2)$ 
  provides 
    $D=4$ conformal algebra and $\hat{o}(5,1)$ can be considered as $D=4$ Euclidean conformal algebra.
     $\hat{o}(3,3)$ algebra 
     is physically more exotic, represents 
      the conformal extension of $\hat{o}(2,2)$ algebra $(\hat{o}(2,2)=\hat{o}(2,1)\oplus\hat{o}(2,1))$, 
       which appears
        as a counterpart of Lorentz algebra 
        in  models with two times  (see e.g. \cite{12},\cite{13}). Further we consider only $\hat{o}(5,1)$ and $\hat{o}(4,2)$ Yang models.}.
\end{enumerate}

The aim of this paper is to consider  the supersymmetric extensions of four Lie algebras $\hat{o}(4,1)$, $\hat{o}(3,2)$, $\hat{o}(5,1)$, $\hat{o}(4,2)$ and  construct respective supersymmetric extensions of Snyder quantum space-times and Yang quantum relativistic phase spaces. For that purpose we will study the superalgebras 
  containing as bosonic subalgebras the spinorial coverings of Lie groups
 describing $D=4$ space-time symmetries  
 which we mentioned above\footnote{The 
   N-component quaternionic real  
   spinors can be described as the pair of 
      N-component complex spinors satisfying symplectic SU(2)-Mayorana subsidiary 
     conditions \cite{14}-\cite{16}.}
\begin{eqnarray}
&\overline{O(4,1)}=U(1,1|H)\simeq USp(2,2), \quad
\overline{O(3,2)}=Sp(4|R)  \label{gr1}
 \\
&\overline{O(5,1)}=SL(2|H)\simeq SU^\star(4), \quad
\overline{O(4,2)}=SU(2,2)\label{gr2}
\end{eqnarray}

The way of constructing NC spaces and NC superspaces from Lie-algebraic and Lie-superalgebraic generators firstly proposed with the use of linear formulae by Snyder \cite{1} and Yang \cite{6}, was further generalized to nonlinear relations by Madore, Gazeau and Buric (see e.g. \cite{16a}-\cite{16c}). Promoting the symmetry generators to quantum space-times or quantum phase space variables we will call "Snyderization procedure". For that purpose let us split the (super)Lie algebra $\hat{g}$ into the (super)Riemanian
 symmetric  pair
\begin{equation}
\hat{g}=\hat{k}\oplus \hat{h}
\end{equation}
where $\hat{g} =\hat{k}\oplus \hat{h}$  and\footnote{Because superalgebra $\hat{g}$ is split into bosonic and fermionic sector $\hat{g}=\hat{g}_{(0)}\oplus\hat{g}_{(1)}$, by $[\cdot,\cdot\rbrace $ we denote graded commutator $(i,j=0,1$ mod $2)$
\begin{equation}\label{gr7}
[g_{(i)},g_{(j)}\}=g_{(i)}g_{(j)}-(-1)^{i\cdot j}g_{(j)}g_{(i)}.
\end{equation}}
\begin{equation}
[\hat{k},\hat{k}\}\subset\hat{h}\qquad [\hat{h},\hat{k}\}\subset\hat{k}\qquad [\hat{k},\hat{h}\}\subset\hat{h}.
\end{equation}
The sub(super)algebra $\hat{h}$ 
   enters as the covariance 
 (super)algebra, and $\hat{o}(3,1) \subset   \hat{h}$\footnote{We will consider here only $D=4$ relativistic-covariant Lorentz Snyder models; if $\hat{o}(4)\subset\hat{h}$ one gets the class of $D=4$ Euclidean Snyder models (see e.g. \cite{19ab}).}; the 
  generators $\hat{k}$ via Snyderization procedure are defining quantum (super) space coordinates, and to quantum phase (super)space if we are able to embedd in $\hat{k}$ some
    quantum-deformation of relativistic Heisenberg algebra.

We observe  that in our Snyderization procedure 
 of superalgebra  $\hat{g}$  all supercharges 
 belonging to $\hat{k}$ 
 will be
    promoted to 
    quantum-deformed spinors,  which will transform under the spinorial covering groups (see e.g. (\ref{gr1})-(\ref{gr2})). In the case of semisimple superalgebras $\hat{g}$ the fermionic odd spinorial generators form an algebraic basis of $\hat{g}$ because for such Lie superalgebras all the bosonic generators can be described as bilinear products of supercharges. Subsequently, in our quantum-deformed covariant superspaces the spinorial quantum coordinates are primary, i.e. they describe the algebraic basis of $\mathcal{U}(\hat{g})$. 
      That 
    property inclined      us to use the names "spinorial Snyder" or "spinorial Yang" models for those obtained by the Snyderization of  supercharges in semisimple Lie superalgebras.

We add that the quantum (super) spaces obtained via described above 
 Snyderization procedure have two important properties:
 
 i) The algebraic construction of quantum (super)spaces inherits from Lie (super)algebras the 
 validity of Jacobi identities i.e. the respective quantum (super)spaces are described by the  associative (super)algebras.
 
 ii) Because underlying classical Lie 
  (super)algebras are endowed with Hopf-algebraic structure, the quantum (super)spaces inherit as well the  coalgebraic  structure, described for classical Lie (super)algebra generators by primitive coproducts.

The plan of this paper is the following. Firstly in Sect. 2 we recall briefly the algebraic construction of quantum 
  D=4 
  Snyder space-time and quantum 
   D=4 Yang relativistic phase spaces. In Sect. 3 we consider two $D=4$ supersymmetric Snyder models of quantum dS and 
   quantum AdS superspaces. Further, in Sect. 4 we describe three types of quantum Yang phase superspaces (one in $D=3$, two in $D=4$), which can be linked with quantum-deformed supersymmetric Heisenberg algebras. In last Section we provide outlook and final remarks.

 It should be mentioned that the
  idea of using Lie (super)algebra relations for description of NC quantum superspaces was already considered earlier in the literature
  (see e.g. \cite{18}-\cite{19}), but these examples studied in the literature were not providing physically the most interesting physically cases of $D=4$ Lorentz-covariant quantum superspaces with AdS and dS quantum fermionic spinors describing their odd sectors. One can add however that interest in Snyder and Yang type models is increasing; in particular recently Zoupanos et al \cite{9},\cite{19a} applied the ideas based on Yang type models in order to describe the dynamics of $D=4$ fundamental interactions, with included quantum gravity sector.

\section{{Snyder quantum space-times and Yang quantum phase spaces}}

\subsection{Snyder quantum space-times}

D=4 dS and AdS algebras are described by the following five-dimensional orthogonal algebras (A=0,1,2,3,4)
\begin{equation}
[M_{AB}, M_{CD}] = i (\eta_{AD} M_{BC} +\eta_{BC}M_{AD} - \eta_{AC}M_{BD} - \eta_{BD}M_{AC})
\label{wz0}
\end{equation}
  with signature $\eta_{AB} = diag(-1,1,1,1,\epsilon)$ and $\epsilon = \eta_{44}=\pm 1$,
 ($\epsilon=1$ for dS algebra and $\epsilon=-1$ for AdS algebra).
 If following the Snyderization procedure we postulate taht $M_{\mu 4} = \frac{1}{\lambda} \hat{x}_\mu$,
  where $\lambda$ is an elementary length in quantum physics given usually by the Planck length
   $\lambda_p = \sqrt{\frac{\hbar G}{c^3}} \simeq 1.6 \cdot 10^{-33}$cm, 
 we obtain besides the Lorentz algebra  generators $M_{\mu\nu}$  ($\mu=0,1,2,3)$  the relations describing NC Snyder 
 space-times\footnote{In the paper we choose $\hslash=c=1$ with only lenght or mass dimensionalities taken into consideration.}
 \begin{eqnarray}
 &
  [ M_{\mu\nu} , \hat{x}_{\rho} ] = i ( \eta_{\nu\rho} \hat{x}_{\mu} - \eta_{\mu \rho} \hat{x}_{\eta})
  \label{wz1}
 \\
 &
 \label{wz2}
  [ \hat{x} _{\mu} , \hat{x}_{\nu} ] = i \epsilon \beta M_{\mu \nu} \qquad \beta = \lambda^{2} > 0
  \label{wz3}
 \end{eqnarray}
 The difference between dS and AdS Snyder 
space-times consists only  in difference of sign on rhs of relation (\ref{wz2}).
The algebra with the basis ($M_{\mu\nu}, \hat{x}_\rho$) introduces algebraically an elementary relativistic quantum system, 
with D=4 quantum (A)dS space-times $\hat{x}_\mu$ as Lorentz algebra module and Lorentz transformations providing the D=4 relativistic covariance
  of  Snyder equations (\ref{wz1}-\ref{wz2}).
Snyder models  (see (\ref{wz0}-\ref{wz2}) are Born-dual ($\hat{x}_\mu \leftrightarrow \hat{p}_\mu$, $M_{\mu\nu}$ 
 unchanged,
$\lambda \to \frac{1}{R}$) to the momentum space realizations $(M_{\mu\nu},\hat{p}_\mu)$ of 
  $\hat{o}(4,1)$ or $\hat{o}(3,2)$ algebras with generators
   describing the automorphisms of five-dimensional pseudospheres 
   \begin{equation}
   x^A \eta_{AB} x^B = - x^2_0 + x^2_1 + x^2_2 +x^2_3 +\epsilon x^2_4 = R^2
   \label{wz23a}
   \end{equation}
    where
 $M_{\mu 4}= R\hat p_{\mu}$ describe the generators of 
curved translations on the pseudospheres $\frac{O(4,1)}{O(3,1)}$ 
or $\frac{O(3,2)}{O(3,1)}$ described by (\ref{wz23a})\footnote{The model realized in fourmomentum space which is Born-dual to Snyder original construction was
 already in sixties applied to the description of dynamics in de-Sitter Universe ($\epsilon=1$), with R describing the 
 cosmological de-Sitter radius (see e.g. \cite{20}-\cite{21}).}.

In fact Snyder constructed his  model as aimed at the description of NC geometry at ultra short (Planckian) distances,
in order to regularize geometrically the ultraviolet divergencies  in renormalization procedure of quantized fields.
Born duality formalizes a physical 
 as well as some philosophical 
 concept that one can relate the micro and macro world  phenomena - the first ones of
quantum nature described by NC geometry, and the second 
  linked with classical de-Sitter dynamics of general relativity
 at very large cosmological distances. 

\subsection{Yang $D=4$ quantum phase spaces}

Already in 1947 C.N. Yang observed that by considering D=6 rotations algebras (i.e. putting in (\ref{wz0})  $A=0,1,2,3,4,5$)
 one can interprete the presence of  rotation generators $M_{5\mu}$ as adding to Snyder model the NC
 fourmomenta $\hat{p} _\mu$. 
 The sixth dimension can be added
   to dS or AdS Snyder model  
  in two-fold way,  by postulating that $\eta_ {55} =\epsilon' = \pm 1$.
 Assuming that $M_{\mu 5}= R \hat{p} _{\mu}$ one gets the following extension of Snyder
 equations (\ref{wz1})-(\ref{wz2}):
 \begin{equation}
 [ M _{\mu\nu}, \hat{p}_{\rho} ] = i (\eta_{\nu\rho} \hat{p}_{\mu} - \eta_{\mu \rho} \hat{p}_{\mu})
 \end{equation}
\begin{equation}\label{2.6}
[ \hat{p}_\mu , \hat{p}_\nu ] = i \epsilon' \gamma  M_{\mu\nu} 
 \qquad   \gamma = \frac{1}{R^2}
\end{equation}
Additionally besides (\ref{wz3}) and (\ref{2.6}) one gets the quantum-deformed canonical phase space commutator
\begin{equation}
[ \hat{x}_{\mu} , \hat{p}_\nu ] = i \frac{\lambda}{R} \eta_{\mu\nu} M_{45} = i \eta_{\mu\nu} \hat{d}
\end{equation}
with operator-valued substitution of Planck constant $\hbar$ by rescaled generator $\hat{d} = \frac{\lambda}{R} M_{45}$, which is a $D=4$ Lorentz scalar
 $( [ M_{\mu\nu}, \hat{d} ] = 0)$. The generator  $M_{45}$ commutes with $\hat{x}_\mu$ and $\hat{p}_\mu$ as follows
 \begin{equation}
 [ \hat{d}, \hat{x}_\mu ] = i \epsilon\lambda R \hat{p}_\mu
 \end{equation}
 
 \begin{equation}
 [ \hat{d} , \hat{p}_\mu ] = i \frac{\epsilon'}{\lambda R} \hat{x} _\mu
 \end{equation}
 and describes in $D=4$ $o(2)$ (if $\epsilon=-\epsilon'$) or $o(1,1)$ (if $\epsilon=\epsilon'$) internal symmetries.
  
 As we mentioned in Introduction we can obtain in such a way four types of Yang models, with two dychotomic parameters
  $\epsilon = \pm 1$ (see(\ref{wz2}))  and $\epsilon' = \pm 1$ (see (\ref{2.6})), which could be also called dSdS, dSAdS, AdSdS, AdSAdS Yang models.

\section{From D=4 SUSY Snyder model to quantum superspaces}

\subsection{Quantum $D=4$ AdS superspace from $D=4$ SUSY AdS Snyder model}
In $\hat{osp}(1|4)$ superalgebra the generators $M_{\mu\nu}$ and $\x_\mu=\lambda M_{\mu 4}$ form the $D=4$ AdS $\hat{o}(2,3)\simeq Sp(4)$ subalgebra, with the generators $M_{\mu\nu}$ describing its $D=4$ Lorentz subalgebra. The $\hat{osp}(1|4)$ superalgebra is obtained by adding to $D=4$ AdS algebra the four additional real Majorana spinor supercharges $Q_\alpha$ which due to "Snyderization procedure" will be 
 interpreted as quantum NC fermionic real components of D=4 AdS spinors
   $\hat{\xi}_\alpha$. The $\hat{osp}(1|4)$ superalgebra is described by the following supersymmetric extension of relations (\ref{wz0})-(\ref{wz2}) with $\beta\rightarrow-\beta$ $(\beta=L^2)$ 
\begin{eqnarray}
 [M_{\mu\nu},M_{\rho\sigma}] &= &
i(\eta_{\mu\sigma}M_{\nu\rho} + \eta_{\nu\rho}M_{\mu\sigma} 
- \eta_{\mu\rho}M_{\nu\sigma} -\eta_{\nu\sigma}M_{\mu\rho})\label{1lineads}\\
\cr
[M_{\mu\nu},\x_{\rho}]&=&i(\eta_{\nu\rho}\x_{\mu}-\eta_{\mu\rho}\x_{\nu})\label{lorr1}\\
\cr 
[\x_{\mu},\x_{\nu}]&=&-i\beta M_{\mu\nu},\\
\cr
 \left\{{\xih}_{\alpha}, \xih_{\beta}\right\}
&= & - 2(C\gamma^{\mu})_{\alpha\beta}\x_{\mu}
  + \beta^{\frac{1}{2}} \left(C\gamma^{\mu\nu}\right)_{\alpha\beta}
M_{\mu\nu}\label{qq}\\
\cr  
 [M_{\mu\nu},\xih_{\alpha}] &=&- {i\over 2} \xih_{\beta}
\left(\gamma_{\mu\nu}\right)^{\beta}_{\ \alpha}\label{lorr2}
\\ \cr
 [\x_{\mu},\xih_{\alpha}] &=&-
 {i\over 2}\beta^{\frac{1}{2}} \xih_{\beta}
\left(\gamma_{\mu}\right)^{\beta}_{\ \alpha}.\label{6lineads}
\end{eqnarray}
where the quantum spinors $\hat{\xi}_\alpha$
 appearing in the place of supercharges 
 have (see(\ref{qq})) the length dimensionality $[\xih_\alpha]=L^{\frac{1}{2}}$. The parameter $\beta$ is the $AdS_4$ Planckian lenght square and $\gamma_\mu$ are $D=4$ Dirac $O(3,1)$ matrices in real Majorana representation;
$
\gamma_{\mu\nu}={1\over 2}(\gamma_\mu\gamma_\nu-\gamma_\nu\gamma_\mu)
$.
Further, by $C_{\alpha\beta}=(\gamma_0)_{\alpha\beta}$ we denote the charge conjugation matrix with the properties
$C^T=-C,~(\gamma^\mu C^{-1})^T=-\gamma^\mu C^{-1}, C^2=-1$

The superalgebra (\ref{1lineads})-(\ref{6lineads}) describes graded associative quantum superspace $(\hat{x}_\mu^{AdS};\hat{\xi}_\alpha)$
\begin{equation}
\mathbb{X}^{(4,4)}_{AdS}=(\hat{x}_\mu^{AdS};\hat{\xi}_\alpha|M_{\mu\nu})
\end{equation}
where the generators $M_{\mu\nu}$ describe the Lorentz covariance algebra (see (\ref{lorr1}) and (\ref{lorr2})). Alternatively, one can Snyderize in $\hat{osp}(1;4)$ only the supercharges, and obtain in such a way purely spinorial model of Snyder type $(\hat{\xi}_\alpha|M_{AB})$, with anticommuting $D=4$ AdS quantized spinors, covariant under the $D=4$ AdS transformations.

\subsection{Quantum $D=4$ dS superspace from $D=4$ SUSY dS Snyder model}

For $D=4$ dS algebra $\hat{o}(4,1)$ the supercharges transform as fundamental spinor realizations of the quaternionic spinorial covering $\overline{O(4,1)}=U(1,1;H)\equiv \hat{osp}(1,2;H)=\hat{usp}(2,2)$ \cite{10}-\cite{12}. The supersymmetrization of $D=4$ dS algebra requires a pair of quaternionic supercharges, which can be equivalently represented by the pair of four-component complex spinors $\hat{Q}_A^i (i=1,2;A=1\dots 4)$ with their quaternionic structure represented by symplectic $SU(2)$ Majorana condition \cite{14}-\cite{16},\cite{26}.

The simple $(N=1)$ $D=4$ dS quaternionic superalgebra is described as the intersection of two complex superalgebras\footnote{We use the notation,
 where N denotes the number of 2-component quaternionic D=4 dS supercharges (see e.g. \cite{14}, \cite{16}).}
\begin{equation}
\label{3.8}
\hat{uu}_\alpha(1,1;1|H)=\hat{su}(2,2;2)\cap \hat{osp}(4;2|C)
\end{equation}
with bosonic sector $\hat{u}(1,1;H)\oplus \hat{u}_\alpha(1;H)\equiv \hat{usp}(2,2)\oplus \hat{o}(2)$.
 Using complex spinors notation the superalgebra (\ref{3.8})
   is described by the following set of (anti)commutators $(A,B=0,1,2,3,4;~\alpha,\beta=1,2,3,4;~ i,j=1,2)$
\begin{eqnarray}
 [M_{AB},M_{CD}] &= &
i(\eta_{AD}M_{BC} + \eta_{BC}M_{AD} 
- \eta_{AC}M_{BD} -\eta_{BD}M_{AC})\label{11lin}\\
\cr
 \left\{\hat{Q}^i_{\alpha}, \hat{Q}^j_{\beta}\right\}
&= & \delta^{ij}(\Sigma_{AB} C)_{\alpha\beta}M^{AB}+\epsilon^{ij}C_{\alpha\beta}T\\
\cr  
 [M_{AB},\hat{Q}^i_{\alpha}] &=&- (\Sigma_{AB})_{\alpha\beta}\hat{Q}_{\beta}^i
\\ \cr
 [M_{AB},T] &=&0\\
 \cr
 [T,\hat{Q}^i_\alpha]&=&-\epsilon^{ij}\hat{Q}^j_\alpha
 \end{eqnarray}
where $\eta_{AB}=diag(1,1,1,1,-1)$, $M_{AB}$ are $\hat{o}(4;1)$ generators, $T$ is a scalar internal $o(2)$ symmetry 
 generator and $\Sigma_{AB}= \frac{1}{2}[ \gamma_A, \gamma_B ]$ represents the 4$\times$4 complex matrix realization of $\hat{o}(4,1)$ algebra (\ref{11lin}).
The complex $\hat{o}(4,1)$ Dirac matrices
      can be chosen for A=0,1,2,3 as real $(\gamma^{dS}_\mu = \gamma_\mu)$
      and the choice for $A=4$ $\gamma^{dS} _4 = i \gamma_5$ is purely imaginary.
     The fermionic supercharger $\hat{Q}^i_\alpha$ satisfy the following quaternionic 
     $SU(2)$-symplectic Majorana condition
\begin{equation}
\hat{Q}^i_\alpha=\epsilon^{ij}(\gamma_5\hat{\bar{Q}}^j)_\alpha,\qquad\Qb=\Q^\dagger C.
\label{jjk}
\end{equation}
In the Snyderization procedure we replace the generators in the coset $\frac{UU_\alpha (1,1;1|H)}{sl(2;\mathbb{C})}$
 by quantum D=4 dS  superspace coordinates  as follows
\begin{equation}
\hat{\psi}_\alpha=\sqrt{i}\beta^{-\frac{1}{4}}\hat{Q}^1_\alpha,\qquad
 \hat{\psi}^\dagger_\alpha=-\sqrt{i}\beta^{-\frac{1}{4}}(\gamma_5\hat{Q}^2)_\alpha,\qquad \x_\mu=\beta^{\frac{1}{2}}M_{\mu4},
\end{equation}
and we obtain 
 the following superalgebra defining quantum D=4 dS superspace
\begin{eqnarray}
 [M_{\mu\nu},M_{\rho\sigma}] &= &
i(\eta_{\mu\sigma}M_{\nu\rho} + \eta_{\nu\rho}M_{\mu\sigma} 
- \eta_{\mu\rho}M_{\nu\sigma} -\eta_{\nu\sigma}M_{\mu\rho})\\
\cr  
[M_{\mu\nu},\x_{\rho}]&=&i(\eta_{\nu\rho}\x_{\mu}-\eta_{\mu\rho}\x_{\nu})\\
\cr 
[\x_{\mu},\x_{\nu}]&=&i\beta M_{\mu\nu}\\
\cr 
 \left\{\hat{\psi}_{\alpha},\hat{ \psi}_{\beta}\right\}
&= & -i(\gamma^{\mu}C)_{\alpha\beta}\x_{\mu}
  +i\beta \left(\gamma^{\mu\nu}C\right)_{\psi\beta}
M_{\mu\nu}\label{srd}\\
\cr 
 \left\{\hat{\psi}^\star_{\alpha}, \hat{\psi}^\star_{\beta}\right\}
&= & - i(\gamma^{\mu}C)_{\alpha\beta}\x_{\mu}
  -i\beta\left(\gamma^{\mu\nu}C\right)_{\alpha\beta}
M_{\mu\nu}\\
\cr 
 \left\{\hat{\psi}_{\alpha}, \hat{\psi}^\star_{\beta}\right\}
&= & -(\gamma_5)_{\alpha\beta}T   \label{ghost}\\
\cr 
 [M_{\mu\nu},\hat{\psi}_{\alpha}] &=&-  
\left(\gamma_{\mu\nu}\right)^{\beta}_{\ \alpha}\hat{\psi}_{\beta}
\\ \cr 
 [\hat{x}_{\mu},\hat{\psi}_{\alpha}] &=&i\beta^\frac{1}{2}
\left(\gamma_{\mu}\right)^{\beta}_{\ \alpha}\hat{\psi}_{\beta}
\\ \cr 
 [T,\hat{\psi}_{\alpha}] &=&\gamma_5 \hat{\psi}^\star_{\alpha}\label{llas}
\end{eqnarray}
with the lenght dimensionalities
\begin{equation}
[M_{\mu\nu}]=0,\qquad[\x_\mu]=1,\qquad
[\hat{\psi}_\alpha]=[\hat{\psi}_\alpha]=\frac{1}{2},\quad[T]=1.
\end{equation}
The superalgebraic relations define the quantum $D=4$ dS superspace
\begin{equation}
\mathbb{X}_{dS}^{(5;4+\bar{4})}=(\hat{x}_\mu^{dS};\hat{\psi}_\alpha,\hat{\psi}_\alpha^\star|M_{\mu\nu},T)
\end{equation}
where $M_{\mu\nu}$ are the Lorentz generators and $T$ describes the internal $O(2)$ symmetries.  generator $T$. 

It follows from relation (\ref{ghost}) and traceless $\gamma_5$ matrix that by putting $\alpha=\beta$ in (\ref{ghost}) one gets $\sum_{\alpha=1}^4|\hat{\psi}_\alpha|^2=0$. The nonvanishing quantum spinors can be therefore only realized in Hilbert-Krein space of states with indefinite metric (see e.g. \cite{26a},\cite{26b}) and local gauging of quaternionic superalgebra (\ref{3.8}) leads to $D=4$ dS supergravity with appearing necessarily gauge ghost fields \cite{27}. 

\section{From $D=4$ $N=2$ SUSY Yang models to quantum phase superspaces}

\subsection{Quantum $D=3$ SUSY AdS phase space from $D=3$ AdS Yang model}
In order to obtain $D=3$ SUSY AdS Yang model to the $\hat{o}(3,2)$ generators $M_{AB}$ (see (\ref{wz0}), $\epsilon=-1$) describing $D=3$ AdS Snyder model we add the pair of real $O(3,2)$ spinorial supercharges $Q^i_\alpha$ $(i=1,2;\alpha=1\dots 4)$ and $\hat{o}(2)$ internal symmetry generator $T$. The underlying $\hat{osp}(2;4)$ superalgebra looks as follows (see e.g. \cite{29})
\begin{eqnarray}
&&\{Q_\alpha^i,Q_\beta^j\}=\delta^{ij}(\gamma^{AB}C)_{\alpha\beta}M_{AB}-\epsilon^{ij}C_{\alpha\beta}T\label{genaa}\\
&&[M_{AB},Q_\alpha^i]=-(\gamma_{AB})_\alpha^{~\beta}Q_\beta^i\\
&&[T,Q_\alpha^i]=-\epsilon^{ij}Q_\alpha^j\label{genaa1}
\end{eqnarray}
where $\gamma_{AB}=\frac{1}{2}[\gamma_A,\gamma_B]$ and $\gamma_A$ denotes the real $O(3,2)$ Dirac-Majorana matrices. Part of the generators in (\ref{genaa})-(\ref{genaa1}) describe the $D=3$ covariance generators
\begin{equation}
M_{rs}\oplus M_{34}\oplus T\simeq \hat{o}(2,1)\oplus \hat{o}(1,1)\oplus \hat{o}(2)
\label{p4.4}
\end{equation}
and the remaining ones are Snyderized as follows
\begin{equation}
M_{3r}\oplus M_{4r}\oplus \tilde{Q}^1_\alpha\oplus \tilde{Q}^2_\alpha\simeq \frac{1}{\lambda}\hat{x}_r\oplus R\hat{p}_r\oplus \hat{\xi}_\alpha\oplus\hat{\pi}_\alpha
\label{p4.5}
\end{equation}
where $(\hat{x}_r,\hat{p}_r)$ and $(\hat{\xi}_\alpha,\hat{\pi}_\alpha)$ describe even and odd canonical pairs of vectorial and spinorial positions and momenta. Denoting $O(1,1)$ generator $M_{45}=\frac{R}{\lambda}\hat{d}$ we obtain the following SUSY-extended quantum-deformed Heisenberg algebra
\begin{itemize}
\item[i)] odd-odd relations
\begin{eqnarray}
&&\{\hat{\xi}_\alpha, \hat{\xi}_\beta\}=\{\hat{\pi}_\alpha, \hat{\pi}_\beta\}=(\gamma^{rs}C)_{\alpha\beta}M_{rs}+\frac{1}{\lambda}(\gamma^{3r}C)_{\alpha\beta}\hat{x}_r\\
&&\qquad\qquad\qquad\qquad\qquad +R(\gamma^{4r}C)_{\alpha\beta}\hat{p}_r+(\gamma^{34}C)_{\alpha\beta}\hat{d}\nonumber\\
&&\{\hat{\pi}_\alpha, \hat{\xi}_\beta\}=C_{\alpha\beta}T\label{a1}
\end{eqnarray}
\item[ii)] even-even relations
\begin{eqnarray}
&&[\hat{x}_r,\hat{x}_s]=\lambda^2M_{rs},\qquad [\hat{p}_r,\hat{p}_s]=\frac{1}{R^2}M_{rs}\nonumber\\
&&[\hat{p}_r,\hat{x}_s]=i\eta_{rs}\hat{d}
\end{eqnarray}
\item[iii)] crossed even-odd relations
\begin{eqnarray}
&&[\hat{x}_r,\hat{\xi}_\alpha]=(\gamma_{3r})_\alpha^{~~\beta}\hat{\xi}_\beta \qquad [\hat{p}_r,\hat{\xi}_\alpha]=(\gamma_{4r})_\alpha^{~~\beta}\hat{\xi}_\beta \nonumber\\
&&[\hat{x}_r,\hat{\pi}_\alpha]=-(\gamma_{3r})_\alpha^{~~\beta}\hat{\pi}_\beta \qquad [\hat{p}_r,\hat{\pi}_\alpha]=-(\gamma_{4r})_\alpha^{~~\beta}\hat{\pi}_\beta.\label{a3}
\end{eqnarray}
\end{itemize}
We see that the relations (\ref{a1})-(\ref{a3}) depend on the quantum superspace coordinates (\ref{p4.5}) as well as the covariance symmetry generators (\ref{p4.4}). In order to solve eq. (\ref{a1})-(\ref{a3}) one can introduce in place of generators (\ref{p4.4}) their irreducible algebraic or fundamental matrix realizations.
\subsection{$D=4$ SUSY AdS phase space from $N=2$ $D=4$ SUSY AdS Yang model}
Using $N=2$ superextension of $D=4$ Minkowskian conformal algebra 
\begin{equation}
\hat{o}(4,2)\simeq \hat{su}(2,2)\xrightarrow{\text{SUSY}}\hat{su}(2,2;2)
\end{equation}
by Snyderization of the coset $\frac{SU(2,2;2)}{SU(2,2)\times U(2)}$ generators one can define the spinorial coordinates  and momenta which span the odd sector of quantum-deformed supersymmetric Heisenberg algebra. The covariance algebra is described by the following extension od $D=4$ Lorentz algebra $(\mu,\nu=0,1,2,3)$
\begin{equation}
M_{\mu\nu}\rightarrow (M_{\mu\nu}, M_{45}, I, I_r)\simeq \hat{o}(3,1)\oplus \hat{o}(1,1)\oplus \hat{o}(2)\oplus \hat{o}(3)\label{4.9}
\end{equation}
where $(I, I_r)$ $(r=1,2)$ describe the internal $u(2)$ symmetries commuting with conformal $o(4,2)$ algebra and the Snyderization procedure should be applied to the remaining generators in order to introduce the quantum-deformed superspace coordinates:
\begin{equation}
(M_{3\mu}, M_{4\mu}, \tilde{Q}_\alpha^a, \tilde{S}_\alpha^a)\xrightarrow{S} (\frac{1}{\lambda}\hat{x}_\mu, R\hat{p}_\mu, \hat{\psi}^a_\alpha, \hat{\pi}^a_\alpha).\label{4.10}
\end{equation}

The fermionic sector of $\hat{su}(2,2;2)$ can be conveniently  described by two pairs of four component real Majorana supercharges $Q^a_\alpha, S_\beta^a$ $(a=1,2;\alpha,\beta=1\dots 4)$, satisfying the following basic superalgebraic relations (we use standard basis $(M_{\mu\nu}, P_\mu, K_\mu, D)$ for the $D=4$ conformal algebra generators; see \cite{24}-\cite{25})
\begin{eqnarray}
&&\{Q_\alpha^a,Q_\beta^b\}=2\delta^{ab}(\gamma^\mu C)_{\alpha\beta}P_\mu\nonumber\\
&&\{S_\alpha^a,S_\beta^b\}=-2\delta^{ab}(\gamma^\mu C)_{\alpha\beta}K_\mu\label{4.11}\\
&&\{Q_\alpha^a,S_\beta^b\}=\delta^{ab}[(\gamma^{\mu\nu} C)_{\alpha\beta}M_{\mu\nu}+2iC_{\alpha\beta}D]\nonumber\\
&&\qquad\qquad\qquad\qquad +\epsilon^{ab}C_{\alpha\beta}I_2+i\tau^{(ab)}_k(\gamma_5C)_{\alpha\beta}I_k\nonumber
\end{eqnarray}
where $k=0,1,2,3$ and $2\times 2$ symmetric matrices $\tau^{(ab)}_k=(1_2,\sigma_1, \sigma_3)$ and generators $(I=I_0, I_1, I_2, I_3)$ describe internal symmetry algebra $u(2)\simeq o(2)\oplus o(3)$. In order to reexpress $P_\mu, K_\mu$ by generators $M_{3\mu}, M_{4\mu}$ we use before employment of the Snyderization procedure the following formulae
\begin{equation}
M_{4\mu}=\frac{1}{\sqrt{2}}(RP_\mu+\frac{1}{\lambda}K_\mu)\xrightarrow{S}\frac{1}{\sqrt{2}\lambda}\hat{x}_\mu
\label{zm1}
\end{equation}
\begin{equation}
M_{5\mu}=\frac{1}{\sqrt{2}}(RP_\mu-\frac{1}{\lambda}K_\mu)\xrightarrow{S}\frac{R}{\sqrt{2}}\hat{p}_\mu.
\end{equation}
The supercharges $\tilde{Q}^a_\alpha, \tilde{S}^a_\alpha$ employed in the Snyderization procedure (see (\ref{4.10})) are defined in terms of supercharges $Q^a_\alpha, S^a_\alpha$ (see (\ref{4.11})) as follows $([Q^a_\alpha]=[\tilde{Q}^a_\alpha])=-\frac{1}{2},[S^a_\alpha]=[\tilde{S}^a_\alpha])=-\frac{1}{2}$
\begin{equation}
\tilde{Q}^a_\alpha=\frac{1}{\sqrt{2}}(Q^a_\alpha+(\lambda R)^{-\frac{1}{2}}S^a_\alpha)\qquad \tilde{S}^a_\alpha=\frac{1}{\sqrt{2}}(S^a_\alpha-(\lambda R)^{\frac{1}{2}}Q^a_\alpha).
\label{zm3}
\end{equation}
The algebra (\ref{4.11}) after using relations (\ref{zm1})-(\ref{zm3}) can be rewritten in term of the supercharger (\ref{zm3}) in the following way
\begin{eqnarray}
&&\{\tilde{Q}_\alpha^a,\tilde{Q}_\beta^b\}=\delta^{ab}[(\gamma^\mu C)_{\alpha\beta}\frac{1}{R}M_{5\mu}+(\gamma^{\mu\nu}C)_{\alpha\beta}M_{\mu\nu}]+\epsilon^{ab}C_{\alpha\beta}I_2\label{ll1}\\
&&\{\tilde{S}_\alpha^a,\tilde{S}_\beta^b\}=-\delta^{ab}[(\gamma^\mu C)_{\alpha\beta}\lambda M_{4\mu}+(\gamma^{\mu\nu}C)_{\alpha\beta}M_{\mu\nu}]-\epsilon^{ab}C_{\alpha\beta}I_2\\
&&\{\tilde{Q}_\alpha^a,\tilde{S}_\beta^b\}=-\delta^{ab}C_{\alpha\beta}D+i\tau^{(ab)}_k(\gamma_5C)_{\alpha\beta}I_k.\label{ll3}
\end{eqnarray}
After Snyderization (see (\ref{4.10})) we obtain the following fermionic odd-odd sector of algebra (\ref{ll1})-(\ref{ll3})
\begin{eqnarray}
&&\{\psi_\alpha^a,\psi_\beta^b\}=\delta^{ab}[(\gamma^\mu C)_{\alpha\beta}\hat{p}_\mu+(\gamma^{\mu\nu}C)_{\alpha\beta}M_{\mu\nu}]+\epsilon^{ab}C_{\alpha\beta}I_2\label{ll1a}\\
&&\{\pi_\alpha^a,\pi_\beta^b\}=-\delta^{ab}[(\gamma^\mu C)_{\alpha\beta}\hat{x}_\mu+(\gamma^{\mu\nu}C)_{\alpha\beta}M_{\mu\nu}]-\epsilon^{ab}C_{\alpha\beta}I_2\\
&&\{\pi_\alpha^a,\psi_\beta^b\}=\delta^{ab}C_{\alpha\beta}D+i\tau^{(ab)}_k(\gamma_5C)_{\alpha\beta}I_k.\label{ll3a}
\end{eqnarray}
We see that in the above relations besides the the "bosonic" and "fermionic" phase space coordinates $(\hat{x}_\mu, \hat{p}_\mu)$,$(\psi_\alpha^a,\pi_\alpha^a)$ enter all the generators of covariance algebra (\ref{4.9}) (we recall that $D=M_{45}$).
\subsection{$D=4$ SUSY dS phase space from $D=4$ SUSY dS Yang model}
Following the supersymmetrization of $\hat{o}(5,1)\simeq sl(2;H)$ algebra, in dS Yang model one should consider the following $N=1$ quaternionic $(N=2$ complex) superalgebras
\begin{equation}
sl(2|H)\simeq su^\star(4)\xrightarrow{\text{SUSY}} sl(2;1|H)\simeq su^\star(4,2)
\end{equation}
with bosonic sector 
\begin{equation}
sl(2|H)\otimes gl(1|H)\simeq su^\star(4)\otimes u^\star(2)
\end{equation}
where $su^\star(4)\simeq\hat{o}(5,1)$ and $u^\star(2)=\hat{o}(2)\oplus\hat{o}(2,1)$. We see that due to quaternionic structure in complex notation we will employ for our $D=4$ Yang models with complex superalgebra $su^\star(4,2)$.

The superalgebras $su^\star(4;2N)$ can be obtained by so called Weyl trick (see e.g. \cite{23}) from the $SU(4;2N)$ superalgebra which supersymetrizes $SU(4)\simeq\hat{o}(6)$. For $SU(4;2N)$ algebra one can introduce the following $Z_4$-grading:
\begin{equation}
\begin{array}{cccc}
L_0 & L_1 & L_2 & L_3 \\
USp(4)\oplus USp(2N) & Q^+ & \frac{SU(4)}{USp(4)}\oplus\frac{U(2N)}{USp(2N)} & Q^-
\end{array}
\label{gens}
\end{equation}
where (see footnote 4)
\begin{equation}
[L_r,L_s\}=L_{r+s}\quad\quad mod~4.\label{multi}
\end{equation}
One passes from $SU(4;2N)$ to $SU^\star(4;2N)$ by multiplication of the generators from sectors (\ref{gens}) according to the following compact formula
\begin{equation}
SU(4;2N)\rightsquigarrow SU^\star(4;2N)\leftrightarrow L_r\rightsquigarrow \exp(\frac{ir\pi}{2})L_r.
\end{equation} 
We choose for $D=4$ SUSY dS phase space the following covariance algebra (compare with (\ref{4.9}))
\begin{equation}
(M_{rs}\oplus M_{45}\oplus I\oplus\tilde{I}_r)\simeq\hat{o}(3,1)\oplus \hat{o}(2)\oplus\hat{o}(2)\oplus \hat{o}(2,1).
\label{internal}
\end{equation}
The remaining generators, which include all supercharges, define via Snyderization procedure the $D=4$ SUSY dS phase superspace
\begin{equation}
(M_{3\mu},M_{4\mu}, z^a_\alpha, u^b_\beta)\xrightarrow{S}(\frac{1}{\lambda}\hat{x}_\mu, R\hat{p}_\mu,\chi^a_\alpha,\rho^a_\alpha)
\label{snyderization}
\end{equation}
where due to quaternionic $SU(2)$-symplectic Majorana condition (see e.g. (\ref{jjk})) it follows that independent degrees of freedom are described by $z_\alpha\equiv z^1_\alpha,\bar{z}_\alpha\sim z^2_\alpha$ and $u_\alpha\equiv u^1_\alpha, \bar{u}_\alpha\sim u^2_\alpha$ and after Snyderization by $\chi_\alpha\equiv\chi_\alpha^1,\bar{\chi}_\alpha\sim\chi^2_\alpha$ and $\rho_\alpha\equiv\rho_\alpha^1,\bar{\rho}_\alpha\sim\rho^2_\alpha$.

The $D=5$ dS superalgebra is the same as $D=4$ Euclidean conformal superalgebra, which has been studied in explicit form (see e.g. \cite{14},\cite{20},\cite{25}). The fermionic odd-odd sector $(A=B=1,2,\dots,5; \bar{\Sigma}_{AB}=\frac{1}{2}[\gamma_A,\gamma_B]C$, where $C$ is a charge conjugation matrix) 

\begin{eqnarray}
&&\{z_\alpha,\bar{u}_\beta\}=2[(\bar{\Sigma}_{AB}\gamma_5)_{\alpha\beta}M^{AB}+C_{\alpha\beta}(\tilde{I}_1+i\tilde{I}_2)]\\
&& \{z_\alpha,\bar{z}_\beta\}=\{u_\alpha,\bar{u}_\beta\}=0\label{ghost}\\
&&\{z_\alpha,u_\beta\}=i(\gamma_5C)_{\alpha\beta}(I+\tilde{I}_3)
\end{eqnarray}
where $I\oplus I_r~(r=1,2,3)$ describe the internal summetry $O(2)\oplus O(2,1)$.

After Snyderization given by formulae (\ref{snyderization}) we get the relations for the complex spinors $\chi_\alpha, \rho_\alpha$ and $\bar{\chi}_\alpha,\bar{\rho}_\alpha$ depending on the covariance algebra generators (\ref{internal}) and bosonic phase space coordinates $\hat{x}_\mu,\hat{p}_\mu$.

It should be mentioned that putting $\alpha=\beta$ in relations (\ref{ghost}) one gets that (see \cite{28}) $\sum_\alpha|z_\alpha|^2=\sum_\alpha|u_\alpha|^2=0$ or $\sum_\alpha|\chi_\alpha|^2=\sum_\alpha|\rho_\alpha|^2=0$ and one can conclude that after quantization, similarly as in Sect. 3.2, the local gauging of superalgebra $su^\star(4;2)$ leads to $D=5$ dS supergravity with ghost gauge fields.

\section{Outlook}

The Snyder dS and AdS models of NC Lorentz-covariant quantum space-time coordinates $\hat{x}_\mu$ are described by the algebras \begin{equation}
  O_{x} (4,1)  \xrightarrow {S}
 \frac{1}{\lambda}{\hat{x}}^{dS}_{\mu}  \oplus M_{\mu\nu}\qquad  O_{x} (3,2) \xrightarrow{S}
  \frac{1}{\lambda} \hat{x}^{AdS}_{\mu} \oplus M_{\mu\nu}  \label{5.1}
 \end{equation}
 Performing semi-dual Born mapp $\hat{x}_\mu \leftrightarrow \hat{p}_\mu,~\lambda\leftrightarrow \frac{1}{R}$
 one gets analogous algebraic structure with Lorentz-covariant quantum NC four-momenta ${\hat{p}}_\mu$

\begin{equation}
O_{p}(4,1)  
 \xrightarrow{S} R\hat{p}_\mu^{dS}  \oplus M_{\mu\nu} \qquad O_{p}(3,2) 
 \xrightarrow{S} R\hat{p}_\mu^{dS}  \oplus M_{\mu\nu}    \label{5.2}
 \end{equation}

Subsequently in dS and AdS Yang models the pairs of algebras (\ref{5.1}), (\ref{5.2}) are embedded in D=6 pseudo-orthogonal quantum deformed Yang algebras
  $\hat{o}_H (5,1)$ and $\hat{o}_H (4,2)$, where subindex $H$ denotes that they contain basic generators $\hat{x}_\mu,\hat{p}_\mu$ defining quantum-deformed Heisenberg algebras. We have the following two diagrams denoting the
   chains of  subalgebras

  \begin{equation}
  \begin{array}{cc}
  \begin{array}{c}
  \hat{o}_{H} (5,1)\ \ 
  \\[-10pt]
  {\begin{array}{c}
 {\rotatebox{-25}{$\Huge\bigcup$}}

  \qquad\qquad 
  \begin{array}{c}
  \phantom{x}
  \\[-4pt]
   {\rotatebox{35}{$\Huge\bigcup$}}
   \end{array}
   \end{array}
   }
  \\
  \hat{o}_x (4,1)  \supset \hat{o}(3,1) \subset \hat{o}_p(4,1)
  \end{array}
   \quad
   \begin{array}{c}
  \hat{o}_{H} (4,2)
  \\[-10pt]
  \begin{array}{c}
 {\rotatebox{-25}{$\Huge\bigcup$}} 
  \qquad\qquad
  \begin{array}{c} \phantom{x}
  \\[-4pt]
   {\rotatebox{30}{$\Huge\bigcup$}}
   \end{array}
   \end{array}
  \\
  \hat{o}_x (3,2)  \supset \hat{o}(3,1) \subset \hat{o}_p(3,2)
  \end{array}
  \end{array}
  \label{5.5}
  \end{equation}
  
  In this paper we introduced in Sect.4 the $N=2$ supersymmetric extensions of $\hat{o}_H (5,1)$ and $\hat{o}_H (4,2)$ algebras, which contain pairs of $N=1$ SUSY Snyder models, with the same (self-dual) Lorentz algebra sectors and dual choices of Lorentz algebra modules.
      
      In the  outlook we would like to comment on some possible directions of future studies, namely:
   \begin{itemize}
   \item[i)]
    Our method of passing from Snyder models describing quantum space-times
       to Yang models  providing quantum-deformed Heisenberg algebra relations is quite general,  
      what also permits to extend supersymmetrically
       Yang models in order to obtain quantum-deformed supersymmetric extension of Heisenberg algebra.
        It should be recalled 
        however the old Snyder idea of adding ``by hand''  to quantum space-time coordinates $\hat{x}_\mu$
       the commuting fourmomenta $p_\mu$.  
        In such an axiomatic method  we postulate further the general covariant formula for the quantum-deformed basic commutator
          (for $\beta$ see (\ref{wz2})) \cite{17}
       
       \begin{equation}
       [ \hat{x}_\mu , p_\nu ] = i  \eta_{\nu \nu} F(\beta p^2) + \beta p_\mu p_\nu G(\beta p^2)
       \label{5.8}
       \end{equation}
       and impose  all required Jacobi identities. Such a problem posed for $D=4$ AdS Snyder models has a general solution with one of the functions $F$,$G$ remaining arbitrary. Such axiomatic approach can be proposed also for supersymmetric extension of Snyder model in order to specify the anticommutator relations for fermionic counterparts $(\hat{\xi}_\alpha,\hat{\pi}_\alpha)$ of the quantum-deformed Heisenberg algebra basis $(\hat{x}_\mu,\hat{p}_\mu)$\footnote{One of the authors (JL) would like to acknowledge that in 2019 using "axiomatic approach" there were also undertaken with E.A.Ivanov the attempts to construct the spinorial version of Snyder model, however without the use of supersymmetry.}.
 
   \item[ii)] If one uses superalgebras for obtaining via Snyderization the spinorial degrees of freedom, the spinors will appear necessarily as Grasmannian, what is desirable in the framework of QFT. One can however employ the "bosonic" coset of matrix groups with employment of spinorial covering  groups - in such a case one gets curved bosonic spinors. A good example is provided by the case of conformal Penrose twistors $(t_A\in\mathbb{C};A=1\cdots 4)$, which are the fundamental representations of $D=4$ conformal "bosonic" group. Describing twistors by bosonic cosets or as $N=1$ superconformal odd cosets one gets the following two different choices (see e.g. \cite{25a})  
  
\begin{equation}
\begin{array}{ccc}
\text{Penrose twistors}~~t_A^{(B)} & & \text{Fermionic twistors}~~t_A^{(F)}\\
t_A^{(B)}=\frac{SU(2,3)}{SU(2,2)\otimes U(1)} & \leftrightarrow & t_A^{(F)} =\frac{SU(2,2;1)}{SU(2,2)\otimes U(1)}.
\end{array}
\end{equation}

\item[iii)] It is known since eighties \cite{27},\cite{28} that local gauging of $D=4$ and $D=5$ quaternionic dS superalgebras leads to the appearance of gauge ghost fields. Recently however it appeared proposals (see e.g. \cite{39}) that $D=4$ dS supergravity without ghost fields can be obtained by suitably spontaneously broken $D=4$ superconformal gravity. Further algebraic understanding of this idea is still desired.

   \end{itemize}

\section*{Acknowledgements}

J.L. and M.W. have been supported by Polish National Science Center, project
2017/27/B/ST2/01902. The main results have been presented at 2021 Corfu Summer Institute, Workshop Sept. 21-27.

\end{document}